%% file: main.tex
\PassOptionsToPackage{hyphens}{url}
\PassOptionsToPackage{inline}{enumitem}

\def\isabridged{1}

\def\isarxiv{1}

\ifdefined\isarxiv{}
	\def\isusenixtemplate{1}
\else
	\def\isasplostemplate{1}%
\fi

\ifdefined\isndsstemplate{}%
	\documentclass[conference]{IEEEtran}
\fi{}
\ifdefined\issigconftemplate{}%
	\documentclass[sigconf,nonacm,natbib=false]{acmart}
\fi{}
\ifdefined\isusenixtemplate{}%
	\documentclass[letterpaper,twocolumn,10pt,natbib=false]{article}
	\usepackage{usenix2019_v3}
\fi{}
\ifdefined\iseurosystemplate{}%
	\documentclass[sigplan,review,twocolumn,anonymous,natbib=false]{acmart}
	\renewcommand\footnotetextcopyrightpermission[1]{}
\fi{}
\ifdefined\isasplostemplate{}%
	\documentclass[pageno]{jpaper}
	
\fi

\usepackage{versions}
\excludeversion{ndss}
\excludeversion{sigconf}
\excludeversion{usenix}
\excludeversion{eurosys}
\excludeversion{asplos}
\ifdefined\isarxiv{}\includeversion{arxiv}\else\includeversion{notarxiv}\fi{}
\ifdefined\isndsstemplate{}\includeversion{ndss}\fi{}
\ifdefined\issigconftemplate{}\includeversion{sigconf}\fi{}
\ifdefined\isusenixtemplate{}\includeversion{usenix}\fi{}
\ifdefined\iseurosystemplate{}{}\includeversion{eurosys}\fi{}
\ifdefined\isasplostemplate{}{}\includeversion{asplos}\fi{}
\ifdefined\isabridged{}%
	\includeversion{Abridged}
	\excludeversion{NotAbridged}
\else
	\excludeversion{Abridged}
	\includeversion{NotAbridged}
\fi
\ifdefined\isanonymous{}%
	\includeversion{Anonymous}
	\excludeversion{NotAnonymous}
\else
	\excludeversion{Anonymous}
	\includeversion{NotAnonymous}
\fi

\usepackage[maxbibnames=99,citestyle=trad-abbrv,eprint=false,doi=false,isbn=false,url=false,backend=biber]{biblatex}

\usepackage[T1]{fontenc}
\usepackage[utf8]{inputenc}
\usepackage{letltxmacro}
\usepackage{todonotes}
\usepackage{booktabs}
\usepackage{amssymb}
\usepackage{amsmath}
\usepackage{xspace}
\usepackage{listings}
\processifversion{ndss}{}%
\usepackage{enumitem}
\usepackage{pifont}%
\usepackage{hyperref}
\usepackage{multirow}
\usepackage{pbox}
\usepackage{color}
\usepackage[acronym]{glossaries}
\usepackage[olditem,oldenum]{paralist}
\usepackage{placeins}
\usepackage[normalem]{ulem}
\usepackage{adjustbox}
\usepackage{colortbl}
\usepackage{diagbox}
\usepackage{cleveref}
\usepackage{csquotes}
\usepackage{algpseudocode}
\usepackage{algorithm}
\usepackage{csvsimple}

\AtEveryBibitem{\clearfield{month}}
\AtEveryBibitem{\clearfield{day}}
\DeclareSourcemap{
  \maps[datatype=bibtex]{
    \map{
      \step[fieldset=eventtitle, null]
      \step[fieldset=publisher, null]
    }
  }
}

\graphicspath{{figures/}}

\definecolor{tagImmWell}{rgb}{0.80,1.0,0.80}
\definecolor{tagImmProt}{rgb}{0.90,1.0,0.90}
\definecolor{tagCorWell}{rgb}{0.90,0.90,0.90}
\definecolor{tagCorProt}{rgb}{0.96,0.96,0.96}

\input{preamble/listings}

\input{preamble/paper-macros}

\newcommand{\LONGNAME}{\protect{Color My World: Deterministic Tagging for Memory Safety}\xspace}
\newcommand{\IMPLNAME}{our \gls{MTE} instrumentation\xspace}
\ifdefined\isarxiv{}%
	\newcommand{\material}{To be released.}
\else
	\newcommand{\material}{\url{https://drive.google.com/drive/folders/14zZbDXRL1JMUoNCn_lUlW8MKmha5dPt7?usp=sharing}}
\fi{}

\glsdisablehyper{}

\setlist*[enumerate,1]{label=\arabic*)}

\setlist[enumerate]{topsep=0pt,itemsep=-1ex,partopsep=1ex,parsep=1ex}
\setlist[itemize]{topsep=0pt,itemsep=-1ex,partopsep=1ex,parsep=1ex}

\makeatletter
\def\blfootnote{\xdef\@thefnmark{}\@footnotetext}
\makeatother

\input{preamble/glossary.tex} %

\begin{document}

\title{\LONGNAME}

\begin{NotAnonymous}
	\processifversion{usenix}{\input{preamble/authors_usenix.tex}}%
	\ifdefined\issigconftemplate{}\input{preamble/authors_sigconf.tex}\fi%
\end{NotAnonymous}

\processifversion{usenix}{\maketitle}%
\ifdefined\isndsstemplate{}\maketitle\fi{}%
\processifversion{asplos}{
	\date{}
	\maketitle
	\thispagestyle{empty}
}

\begin{abstract}
\input{sections/000.abstract}

\end{abstract}

\ifdefined\issigconftemplate{}\maketitle\fi{}%
\ifdefined\iseurosystemplate{}\maketitle\fi{}%

\begin{NotAnonymous}
	\ifdefined\isndsstemplate{}\input{preamble/authors_ndss}\fi{}%
	\ifdefined\issigconftemplate{}\input{preamble/authors_sigconf}\fi{}%
\end{NotAnonymous}

\input{sections/010.introduction}

\input{sections/020.background}

\input{sections/030.model}
\input{sections/040.requirements}

\input{sections/050.design}

\input{sections/060.analysis}

\input{sections/070.implementation}

\input{sections/080.evaluation}

\input{sections/090.related}

\input{sections/100.discussion}

\processifversion{NotAnonymous}{\input{sections/110.acknowledgments}}

\printbibliography{}

\end{document}

%% file: preamble/listings.tex
\definecolor{lst:comment}{rgb}{0.0,0.0,0.0}
\definecolor{lst:instructions}{rgb}{0.4,0.1,0.1}
\definecolor{lst:registers}{rgb}{0.1,0.4,0.1}
\definecolor{lst:bg}{rgb}{0.98,0.98,0.98}
\lstdefinelanguage{aarch64}{
    keywords={pacga,stp,mov,bl,ldp,cmp,jnz,ret,str,ldr,autia,pacia,eor,autib,pacib,b,cbnz,cbz,tst,beq,stg,subs,st2g},
    keywordstyle=\bfseries\color{lst:instructions},
    keywords=[2]{LR, CR, Xd, Xr, SP,XZR, X28},
    keywordstyle=[2]\color{lst:registers},
    breaklines=true,
    morecomment=[l]{;},
    commentstyle=\itshape\color{lst:comment},
}

%% file: preamble/paper-macros.tex
\LetLtxMacro{\todonote}{\todo}
\renewcommand{\todo}[2][]
{\todonote[inline, caption={#2}, size=\footnotesize, #1]
{\renewcommand{\baselinestretch}{0.5}\selectfont#2\par}}

\makeatletter
\newcommand\newtag[2]{#1\def\@currentlabel{#1}\label{#2}}
\makeatother

\newcommand\taggedparagraph[1]{\noindent\textbf{#1.}}

\newcommand{\inlinec}[1]{\texttt{#1}\xspace}
\newcommand{\inlinellvm}[1]{\texttt{#1}\xspace}
\newcommand{\inlineasm}[1]{\texttt{\lowercase{#1}}\xspace}
\newcommand{\instruction}[1]{\texttt{\MakeLowercase{#1}}\xspace}

\newcommand{\dOne}{\ding{182}\xspace}\newcommand{\dTwo}{\ding{183}\xspace}\newcommand{\dThree}{\ding{184}\xspace}
\newcommand{\dFour}{\ding{185}\xspace}
\newcommand{\dSeven}{\ding{188}\xspace}
\newcommand{\dCOne}{\ding{192}\xspace}\newcommand{\dCTwo}{\ding{193}\xspace}\newcommand{\dCThree}{\ding{194}\xspace}
\newcommand{\dCFour}{\ding{195}\xspace}

\newcommand{\theAttacker}{\ensuremath{\mathcal{A}}\xspace}

\newcommand{\StackSafetyAnalysis}{\texttt{StackSafetyAnalysis}\xspace}
\newcommand{\MTStack}{\inlinec{MTStack}}
\newcommand{\MTStackLoad}{\inlinec{MTStackLoad}}

\newcommand{\UseInfo}{\inlinec{UseInfo}}

\newcommand{\guarded}{guarded\xspace}

\newcommand{\safeTag}{\ensuremath{\mathtt{0b1100}}\xspace}

\newcommand{\ptrUnsafeTag}{\ensuremath{\mathtt{0b10xx}}\xspace}
\newcommand{\unsafeTag}{\ensuremath{\mathtt{0b0xxx}}\xspace}

\newcommand{\LLVMversion}{LLVM 12\xspace}
\newcommand{\qemuVersion}{QEMU 6.0.0\xspace}
\newcommand{\benchmarkHardware}{Kunpeng 920\xspace}

\newcommand{\TagProt}{Tag forgery prevention\xspace}
\newcommand{\tagProt}{tag forgery prevention\xspace}

\newcommand{\safeColumn}{\emph{Safe}\xspace}

\newcommand{\specPerfOverhead}{\ensuremath{13.6\%}\xspace}
\newcommand{\specPerfOverheadMax}{\ensuremath{26.8\%}\xspace}
\newcommand{\specMemOverhead}{\ensuremath{19.3\%}\xspace}
\newcommand{\specSizeOverhead}{\ensuremath{21.7\%}\xspace}

\newcommand\thefont{\expandafter\string\the\font}

%% file: preamble/glossary.tex
\newacronym{ABI}{ABI}{application binary interface}
\newacronym{ADI}{ADI}{Application Data Integrity}
\newacronym{ASLR}{ASLR}{address-space layout randomization}
\newacronym{ASAN}{ASAN}{AddressSanitizer}
\newacronym{BTI}{BTI}{Branch Target Indicator}
\newacronym{CFG}{CFG}{control-flow graph}
\newacronym{CFI}{CFI}{control-flow integrity}
\newacronym{DEP}{DEP}{data-execution prevention}
\newacronym{DFI}{DFI}{data-flow integrity}
\newacronym{DOP}{DOP}{data-oriented programming}
\newacronym{EL}{EL}{exception level}
\newacronym{FVP}{FVP}{Fixed Virtual Platform}
\newacronym{GEP}{\inlinellvm{GEP}}{\inlinellvm{getelementptr}}
\newacronym{GM}{geo. mean}{geometric mean}
\newacronym{HWASAN}{HWASAN}{hardware-assisted AddressSanitizer}
\newacronym{IR}{IR}{intermediate representation}
\newacronym{ISA}{ISA}{instruction set architecture}
\newacronym{LTO}{LTO}{link-time optimization}
\newacronym{MAC}{MAC}{message authentication code}
\newacronym{MMU}{MMU}{memory management unit}
\newacronym{MTE}{MTE}{Memory Tagging Extension}
\newacronym{MTSSA}{MTSSA}{memory tagging aware stack safety analysis}
\newacronym{PA}{PA}{Pointer Authentication}
\newacronym{PAC}{PAC}{pointer authentication code}
\newacronym{PEI}{\inlinec{PEI}}{\inlinec{PrologEpilogInserter}}
\newacronym{ROP}{ROP}{return-oriented programming}
\newacronym{SoC}{SoC}{system-on-a-chip}
\newacronym{SSA}{SSA}{static single assignment}
\newacronym{SP}{SP}{stack pointer}
\newacronym{TBI}{TBI}{top-byte ignore}
\newacronym{TOCTOU}{TOCTOU}{time-of-check-time-of-use}
\newacronym{VA}{VA}{virtual address}
\newacronym{VLA}{VLA}{variable-length array}

%% file: preamble/authors_usenix.tex
\begin{NotAnonymous}

\author{%
\rm Hans Liljestrand\\
\normalsize{Univresity of Waterloo, Canada}\\
\normalsize{hans@liljestrand.dev}
\and
\rm Carlos Chinea\\
\normalsize{Huawei Technologies Oy, Finland}\\
\normalsize{carlos.chinea.perez@huawei.com}
\and
\rm R\'emi Denis-Courmont\\
\normalsize{Huawei Technologies Oy, Finland}\\
\normalsize{remi.denis.courmont@huawei.com}
\and
\rm Jan-Erik Ekberg\\
\normalsize{Huawei Technologies Oy, Finland}\\
\normalsize{Aalto University, Finland}\\
\normalsize{jan.erik.ekberg@huawei.com}\\
\and
\rm N. Asokan\\
\normalsize{University of Waterloo, Canada}\\
\normalsize{Aalto University, Finland}\\
\normalsize{asokan@acm.org}
}

\end{NotAnonymous}

%% file: sections/000.abstract.tex
Hardware-assisted memory protection features are increasingly being deployed in COTS processors. %
ARMv8.5 Memory Tagging Extensions (MTE) is a recent example, which has been used to provide probabilistic checks for memory safety. This use of MTE is not secure against the standard adversary with arbitrary read/write access to memory. Consequently MTE is primarily used as a software development tool. 

We present the first design for deterministic memory protection using MTE that can resist the standard adversary, and hence is suitable for \textbf{post-deployment memory safety}. Using static analysis we compartmentalize memory allocations into safety classes, and describe several compile-time instrumentation techniques to efficiently shield data categorized as ``safe'', such as function returns, from adversaries who exploit memory errors in the code.  We incorporate our design into LLVM Clang, implementing static analysis and subsequent MTE instrumentation.
Via a comprehensive evaluation we show that our scheme is effective and achieves a low run-time overhead of \specPerfOverhead and code-size overhead of \specSizeOverhead (\glsdesc{GM}).

%% file: sections/010.introduction.tex
\section{Introduction}%
\label{sec:intro}

Hardware-assisted memory tagging, originally introduced in early computer architectures~\cite{mayerArchitecture1982} but largely abandoned shortly thereafter, has resurfaced in recent processor architectures such as lowRISC~\cite{bradburyTagged2014}, SPARC M7~\cite{aingaran2015m7}, and ARMv8.5-A~\cite{arm-mte8.5,gretton-dannArm2018blog}.
It is a powerful technique that allows detection of memory errors using a lock-and-key mechanism where both pointers and memory allocations are associated with \emph{tags}; a pointer is allowed to access a chunk of memory if and only if both the pointer and the memory allocation have matching tag values (``colors'').
Tools like \gls{HWASAN}~\cite{clang-11-hwasan} use a randomized scheme to tag each memory allocation (and corresponding pointers) with a random tag. They can then efficiently detect many spatial and temporal memory access violations. Since the tags are randomly assigned, these tools can only provide a \emph{probablistic} guarantee of correctness. Furthermore, the guarantee is \emph{limited} because the tags are short (a few bits long). However, these tools are typically used for testing in the software development phase. In such benign, non-adversarial settings, limited correctness guarantees are sufficient.

A natural question is whether and how hardware-assisted memory tagging can be used for ensuring run-time memory safety \emph{after deployment}.
Unlike the software development phase, post-deployment is a potentially adversarial environment: detecting memory-safety violations after deployment therefore must be robust in the presence of an intelligent adversary. The standard adversary model used in memory safety literature assumes an adversary that can exploit a memory vulnerability to \emph{read from or write to arbitrary memory locations}.
Memory tagging schemes that use randomized tag assignment cannot withstand such an adversary that can (a) learn tag values and subsequently inject bogus pointers with correct tags, or (b) repeat an attack until injected pointer tags match by chance. Therefore, existing memory tagging schemes, with their limited probabilistic guarantees, cannot be used directly to ensure post-deployment memory safety.

In this paper, we describe the \emph{first} hardware-assisted memory tagging approach that enforces post-deployment run-time memory protection even in the presence of the standard (strong but realistic) adversary that can forge tags and repeat an attack indefinitely.
Unlike prior schemes, we target \emph{deterministic} memory protection without using probabilistic random tagging.
To achieve this, we first use static analysis to classify memory allocations into different classes based on their level of \emph{operational safety}.
But because the memory tag of a pointer stored in memory is embedded within the pointer itself, the adversary can overwrite it freely.
Consequently, we use a \emph{\tagProt} scheme to prevent attacker-controlled pointers from accessing safe allocations.
Specifically, we can guarantee the integrity of provably safe allocations even in the presence of memory errors that affect the less secure classes.
Furthermore, our static analysis is \emph{memory-tagging aware}: it can detect allocations that are safe \emph{only when coupled with our tagging scheme} or other similar run-time enforcement.

We realize our scheme using ARMv8.5 \gls{MTE}~\cite{arm-mte8.5} which is slated to be deployed soon~\cite{samsungSamsung2022}.
Our implementation is based on the LLVM compiler framework and covers both our static analysis and instrumentation.

We claim the following contributions:

\begin{itemize}

    \item The \textbf{first design for deterministic memory protection using hardware-assisted memory tagging}, guaranteeing data integrity for safe classes, even against the \textbf{standard adversary} that can break prior memory-tagging schemes. Our scheme can thus be used to ensure \textbf{post-deployment integrity of protected data}. (Section~\ref{sec:design})
    
    \item A complete ARMv8.5-A \gls{MTE} based implementation of our design built on the \LLVMversion{} compiler framework,  including augmented stack safety analysis (Section~\ref{sec:analysis}), and compiler back-end modifications (Section~\ref{sec:implementation}).  We will open-source our implementation\footnote{\material}. 

    \item A comprehensive evaluation including security analysis (Section~\ref{sec:security-evaluation}), functional evaluation  using QEMU with \gls{MTE} support (Section~\ref{sec:functional-evaluation}), and performance analysis using SPEC CPU 2017 indicating a performance overhead of \specPerfOverhead and code-size overhead of \specSizeOverhead(\glsdesc{GM}) (Section~\ref{sec:performance-evaluation}), shows the effectiveness of our scheme. 
    
\end{itemize}

%% file: sections/020.background.tex
\section{Background}

\subsection{ARMv8.5-A Memory Tagging Extension}%
\label{sec:arm-mte}%
\label{sec:background-mte}

\begin{figure}
  \centering
  \includegraphics[width=1\columnwidth]{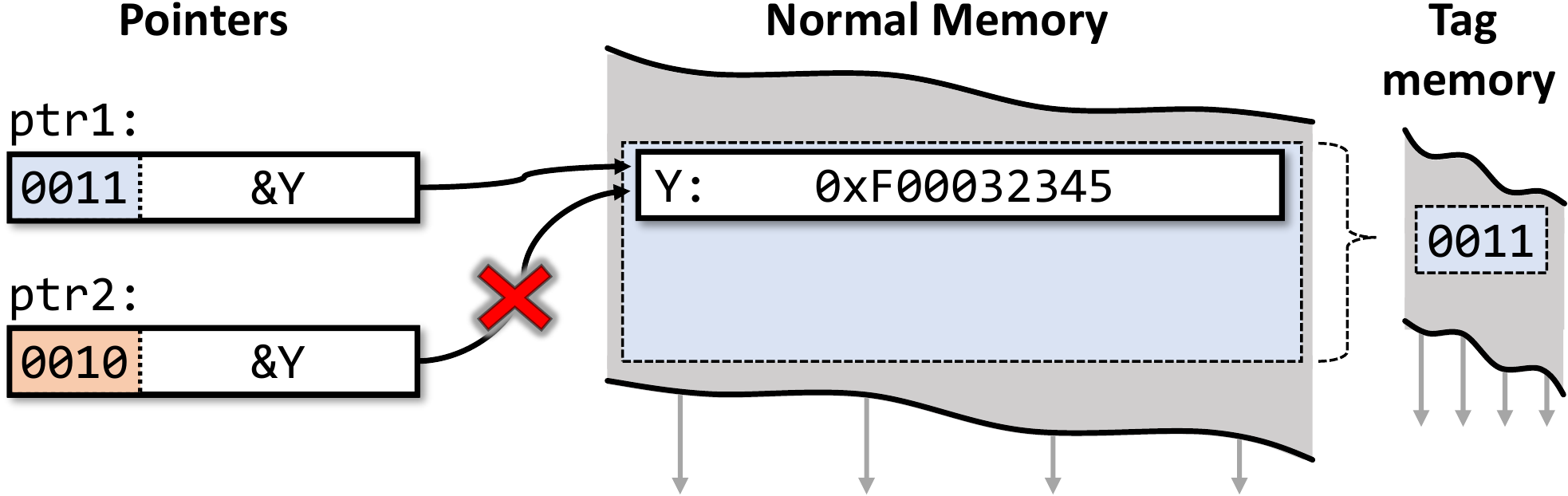}
  \caption{
    \Gls{MTE} supports ``coloring'' memory and pointers with \emph{allocation} and \emph{address tags}, respectively.
    \processifversion{NotAbridged}{%
        Memory accesses must then use a pointer with the address tag matching the allocation tag of the accessed memory.
        In this example, \texttt{ptr2} is denied access to \texttt{Y} because its address tag does not match the allocation tag of the addressed memory.
    }
  }\label{fig:tagsinaction}
\end{figure}

The ARMv8.5 \gls{ISA} introduces the \glsfirst{MTE} feature, a hardware primitive for \emph{memory tagging} (colloquially referred to as \emph{memory coloring})~\cite{arm-mte8.5,gretton-dannArm2018blog}.
\Gls{MTE} allows 4-bit tags to be assigned to each memory allocation and address (at 16-byte granularity). 
These are referred to as \emph{allocation tags}  and \emph{address tags} respectively.
An \emph{allocation tag} is a 4-bit value associated with every \emph{memory granule} (a 16-byte region of memory).
Allocation tags are stored in hardware-protected \emph{tag memory}.
An \emph{address tag} is a 4-bit value stored within the highest-order byte of a pointer.
\begin{NotAbridged}
ARM-A processors have a \gls{TBI} feature to instruct the memory management hardware to ignore the topmost byte during address translation. In conjunction, \gls{MTE} uses the lower half to store the tag value. Allocation tags are tracked by the hardware. During any memory access, the allocation tag of the accessed memory is compared to the address tag of the pointer attempting access (\Cref{fig:tagsinaction}).
Memory accesses that use incorrect tags will trap to the operating system, allowing it to choose a course of action. 
\end{NotAbridged}

Programs can be instrumented with explicit \gls{MTE}-specific instructions to tag memory and associated pointers according to the tagging strategy.
\gls{MTE} provides instructions for writing and reading allocation tags in memory, as well as for efficiently manipulating the address tag bits of a pointer in a register. 
\begin{NotAbridged}
The \instruction{ADDG} and \instruction{SUBG} instructions allow a pointer in a register and its tag to be generated based on a reference pointer (and tag). 
\instruction{STG} tags a granule of memory, and \instruction{LDG} allows reading the tag of a specific memory location. 
In total, \gls{MTE} adds 15 new instructions for operating on the \gls{MTE} tagging framework.
\end{NotAbridged}

For efficiency, \gls{MTE} also allows \emph{unchecked accesses} in some cases.
First, no accesses via the stack pointer, or the stack pointer with an immediate offset, are checked.
This allows compiler-generated code that modifies the stack---e.g., to store local variables in memory---to avoid tag checks.
\Gls{MTE} also supports unchecked access using a wildcard tag.
For user space, the wildcard tag is \texttt{0b0000} and is enabled via the \inlineasm{TCR\_ELx.TCMA} configuration register.
If enabled, any pointer with the tag \texttt{0b0000} will not undergo any tag checks.
\begin{NotAbridged}
\Gls{MTE} can be used in two modes: a synchronous mode where mismatching tags immediately trap, or an asynchronous mode that only sets a flag to indicate that a mismatch has been encountered at some point during execution. The former allows for precise error handling, whereas the latter is more performant but does not facilitate precise debugging.
\end{NotAbridged}

Memory tagging can be used to detect spatial and temporal memory errors.
For example, use-after-free errors for heap allocations (or use-after-return for stack allocations) can be protected against by assigning random tags to newly allocated memory regions and clearing the tags on freeing memory~\cite{serebryanyARM2019}. 
Re-allocated memory will, with probability $\frac{15}{16}$, be assigned a different tag, which prevents access by any remaining dangling pointers.
However, because the address tag is stored within the pointer itself, such solutions do not lend themselves to post-deployment run-time protection against an adversary that can overwrite pointers and their address tags.

\subsection{LLVM memory tagging on ARM}%
\label{sec:background-llvm-memory-tagging}

At the time of writing LLVM provides two sanitizers that utilize ARM \gls{MTE}: the Clang \gls{HWASAN}~\cite{clang-11-hwasan} and LLVM MemTagSanitizer~\cite{llvm-11-memtagsan}.
The former, on ARM architectures, uses the \gls{TBI} feature to reserve the most-significant byte of pointers for a software-defined tag that is checked by software-only instrumentation that largely targets testing use cases.
In contrast, MemTagSanitizer is designed to use \gls{MTE} for protecting programs both during testing as well for during post-deployment run-time protection.
On function entry, it selects a random tag for the first tagged frame slot, and then tags subsequent slots with incrementing tags.
This provides probabilistic detection of arbitrary memory errors, and can prevent some, but not all, linear buffer overflows.
Neither HWASAN or MemTagSanitizer considers an adversary that has arbitrary memory-read access, and therefore, can reliably forge tags.
In this work, we use the standard adversary model typically used in memory protection~\cite{szekeresSoK2013}, and specifically consider an adversary that has these abilities.

\subsection{LLVM \StackSafetyAnalysis}%
\label{sec:llvm-ssa}

LLVM \StackSafetyAnalysis~\cite{llvm-11-stacksafetyanalysis} detects stack-based variables that are ``safe'', i.e., cannot cause memory errors through their use.
It was originally introduced to support the Clang SafeStack~\cite{clang-11-safestack} used for code-pointer integrity~\cite{kuznetsovCodepointer2014}, but can also be used to optimize approaches such as HWASAN\@.
For each stack-based allocation, the analysis indicates the smallest memory region that contains all accesses by any direct or derived pointer based on the allocation.
If the analysis is inconclusive, it indicates that the full memory range is accessed.
Moreover, \StackSafetyAnalysis does not track pointers written to memory, and instead assumes that such pointers can be corrupted or otherwise misused. 
Without full memory safety or fault isolation, this is the only safe assumption.

\begin{NotAbridged}
  The \StackSafetyAnalysis{} implementation is split into two passes.
  The first \textit{intra-procedural} pass analyzes the use of stack-based allocations and pointer-type function arguments within the scope of one function.
  The access range is always relative to the \emph{base pointer}, i.e., the start of the allocation or the initial value of a function argument pointer.
  A second \textit{inter-procedural} analysis combines the local analysis to facilitate tracking of pointers passed between functions. 
  When pointers are passed as arguments, the offset to the base pointer is used to adjust the argument access range before the ranges are merged.
  As noted above, pointer tracking is not done through memory and will conservatively mark untracked pointers as accessing the full set of memory addresses.
\end{NotAbridged}

%% file: sections/030.model.tex
\section{Adversary Model}%
\label{sec:model}

In this work, we consider a powerful adversary, \emph{\theAttacker}, that can exploit a memory vulnerability to overwrite program data, including pointers and their address tags.
We assume that the memory is protected with a W$\oplus$X policy.
Thus, \theAttacker can neither modify program code nor execute data (e.g., inject and run code on the program stack), but can read all process memory, including the allocation tags used by \gls{MTE}.
For user-space protection, we assume that the operating system is trusted, and that \theAttacker cannot modify \gls{MTE} configuration.
This adversary model is consistent with prior work on memory safety~\cite{szekeresSoK2013}.

The ability to read arbitrary memory allows \theAttacker to determine all memory tags used by the process.
This renders probabilistic tagging schemes ineffective against \theAttacker that can also inject arbitrary pointers and tags.
However, even without read access to allocation tags, given the limited range from which tag values are drawn, \theAttacker can repeat an attack until it succeeds.
\begin{NotAbridged}
  We assume \theAttacker can repeatedly relaunch and retry an attack.
\end{NotAbridged}

%% file: sections/040.requirements.tex
\section{Goals and Requirements}%
\label{sec:requirements}

Our goal is to partition allocations into protection domains based on their memory-safety properties, use memory tagging to enforce domain separation, and where possible, enforce in-domain memory integrity (e.g., by preventing overflows).
The former is similar to the \emph{safe stack} employed by Kuznetsov \emph{et al.}~\cite{kuznetsovCodepointer2014} that aims to isolate provably memory-safe data from other data (such as buffers that might overflow).
In particular, we use the notion of \emph{memory-safe allocations}.
Informally, a memory-safe allocation is one where all pointers \emph{based on} the allocation are safe, and cannot cause overflows or other memory errors.
See \Cref{sec:design} for a more detailed definition.

We define the following requirements for our solution:

\begin{enumerate}[label=\textbf{{R\arabic*}},ref=\textbf{{R\arabic*}}]
\itemsep0em
	\item\label{req:safe}\textit{Isolate safe allocations}: Allow access to a safe allocation only through legitimate pointers to it.
  \item\label{req:compart}\textit{Isolate unsafe allocations}: Prevent unsafe memory accesses from compromising data in safe allocations.
  \item\label{req:forgery}\textit{Resist address tag forgery}: Prevent \theAttacker from accessing safe allocations using forged pointers.
  \item\label{req:overflow}\textit{Resist buffer overflow}: Prevent \theAttacker from exploiting linear overflows to violate allocation bounds.
  \item\label{req:disclosure}\textit{Resist memory disclosure}: Remain effective if \theAttacker can read the entire process address space, including tags.
  \item\label{req:comp}\textit{Be compatible}: Be applicable to typical C code, without source code modifications.
  \item\label{req:perf}\textit{Be efficient}: Impose only minimal run-time performance and memory overhead, while meeting \ref{req:compart}--\ref{req:comp}.
\end{enumerate}

%% file: sections/050.design.tex
\section{Design}%
\label{sec:design}\label{sec:memory-safety-def}

\begin{figure}[tp]
\centering
\includegraphics[width=0.6\columnwidth]{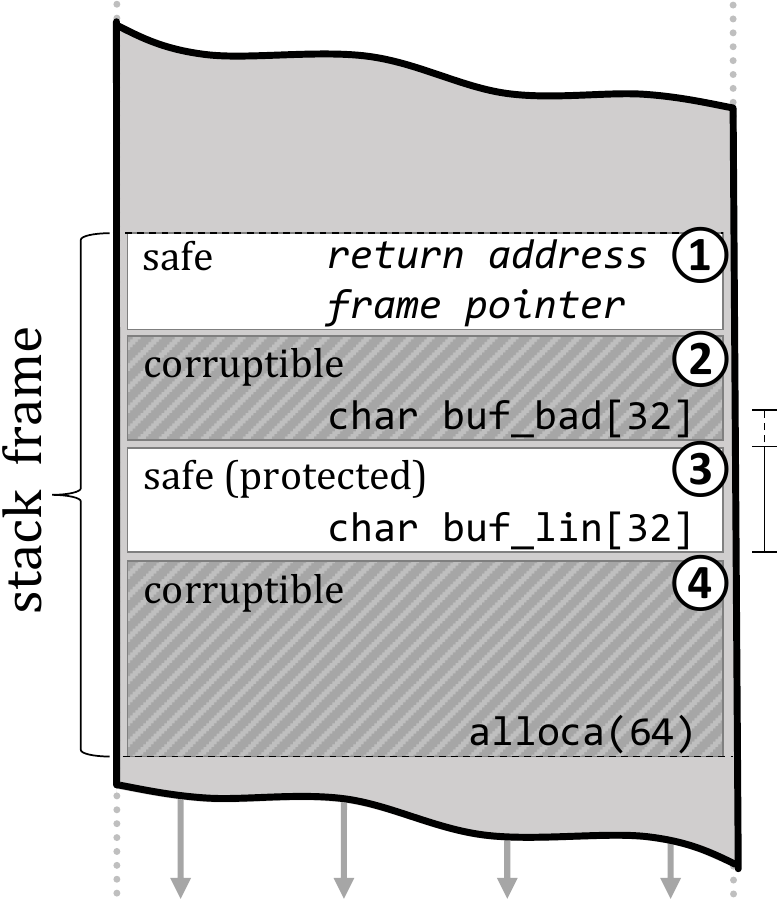}
\caption{
	We tag allocations based on their safety class.
	\processifversion{NotAbridged}{%
        Linear overflows are thwarted by adjusting allocation layout and by adding tagged memory guards.
    }
	\Cref{lst:code-example} show the corresponding function.
}\label{fig:stackframe}
\end{figure}

Our design utilizes memory tagging without relying on random or hidden tags (\ref{req:disclosure}), instead we:
\begin{enumerate*}
\item \emph{isolate} memory allocations into distinct protection domains (\ref{req:safe}, \ref{req:compart}),
\item \emph{alternate} memory tags to prevent linear overflows (\ref{req:overflow}), and
\item \emph{prevent address tag forgery} by explicitly setting tag bits of pointers that could be controlled by \theAttacker (\ref{req:forgery}).
\end{enumerate*}
We leverage an \gls{MTE}-aware static analysis (\Cref{sec:analysis}) to classify memory allocations into two distinct protection domains based on whether we can guarantee the allocations to be \emph{memory safe}.

We define an allocation as \emph{memory safe if all dereferences of pointers \emph{based on} the allocation are either \emph{safe}, or cause the program to terminate.} 
Our definition of memory-safe allocations is based on Kuznetsov \emph{et al.}, who define a \emph{safe dereference} as one that only accesses the memory object that the pointer is \emph{based on}, and \emph{based on} as:
``\emph{a pointer is \emph{based on} a target object X iff the pointer is obtained at runtime by (i) allocating X on the heap, (ii) explicitly taking the address of X, if X is allocated statically, such as a local or global variable, or is a control flow target (including return locations, whose addresses are implicitly taken and stored on the stack when calling a function), (iii) taking the address of a sub-object y of X (e.g., a field in the X struct), or (iv) computing a pointer expression (e.g., pointer arithmetic, array indexing, or simply copying a pointer) involving operands that are either themselves based on object X or are not pointers}~\cite{kuznetsovCodepointer2014}.''
For brevity, we will use \emph{safe} and \emph{unsafe} as shorthand to refer to memory safe and memory unsafe, respectively.

\begin{lstlisting}[language=c,float=,floatplacement=tp,caption={
  Code example corresponding to \Cref{fig:stackframe}.
  \processifversion{NotAbridged}{%
    \inlinec{buf\_lin} belongs to the \guarded category, as it is only used in a loop with a known starting point and indexing step (\Cref{line:code-example:loop}).
    \inlinec{buf\_bad} is unsafe because the index \inlinec{end} is unknown at \Cref{line:code-example:bad-indexing}.
    The \inlinec{alloca} on \Cref{line:code-example:alloca-to-global} leaks the pointer to a global variable, and so, is treated as unknown.
  }
  },label=lst:code-example,linerange={3-15}]
void func(unsigned end) {
    char buf_bad[32]; /* \label{line:code-example:bad} */
    char buf_lin[32];

    g_ptr = alloca(64); /* \label{line:code-example:alloca-to-global} */

    buf_bad[end] = 0; /* \label{line:code-example:bad-indexing} */

    for (unsigned i = 0; i < end; ++i) /* \label{line:code-example:loop} */
        buf_lin[i] = i;

    // ...
\end{lstlisting}

If all pointer dereferences in a program are safe, then the program as a whole is memory safe and each allocation will only be modified by pointers that are based on it.
Note that without program-wide memory safety, a safe allocation can still be corrupted by unsafe pointer dereferences; either using pointers based on unsafe allocations or pointers injected by \theAttacker.
However, using our \gls{MTE} tagging scheme and \emph{\tagProt} (explained in \Cref{sec:design-tag-prot}), we can prevent such corruption (\ref{req:forgery}) and protect safe allocations without requiring full, program-wide, memory safety.

We divide safe allocations into three types:
\begin{enumerate*}
  \item \emph{implicitly-safe} allocations that are only dereferenced by compiler-generated code and are thus always safe,
  \item \emph{provably-safe} allocations for which our analysis can conclusively prove that all dereferences of pointers based on the allocation are safe, and
  \item \emph{\guarded} allocations that cannot be proven safe, but for which we can prove that all unsafe dereferences of pointers based on the allocation are prevented by our instrumentation.
\end{enumerate*}

Our approaches to protect safe allocations is thus three-fold:
\begin{enumerate*}
  \item we use static analysis (\Cref{sec:analysis}) to identify provably-safe and \guarded allocations by verifying that all dereferences of pointers based on those allocations are safe or can be \guarded;
  \item we use \gls{MTE} to prevent unsafe dereferences using pointers based on \guarded allocations (\Cref{sec:design-guarded}) from corrupting other allocations; and
  \item we apply \gls{MTE} to isolate unsafe allocations and \tagProt (\Cref{sec:design-tag-prot}) to ensure that neither corrupted pointers nor attacker-injected pointers can dereference safe allocations.
\end{enumerate*}
\Cref{fig:stackframe} and \Cref{lst:code-example} illustrate the different types of allocations and will be used as a running example to explain our approach.

To maximize the set of safe allocations we also track pointers when they are stored in memory.
Our intent is to guarantee that pointers based on safe allocations cannot be corrupted in memory or be used in unsafe dereferences.
Consequently, if a such a pointer is ever stored in a non-safe allocation, we also mark the allocation it is based on as unsafe.
All other pointers stored to or loaded from memory are assumed to be corrupted or used unsafely.
Because our definition of memory safety is allocation-based we must separately address the problem of narrowing~\cite{gilThere2018} with respect to pointers.
Consequently, we also verify that pointers stored in compound structures\processifversion{NotAbridged}{---e.g., \inlinec{ptr} in \Cref{lst:allocation-bound-problem}---} cannot be corrupted, or else we use \tagProt to ensure that such pointers always point to unsafe allocations.

\subsection{Isolating unsafe allocations and pointers}%
\label{sec:design-fault-isolation}%
\label{sec:design-tag-prot}

We isolate unsafe allocations using ARM \gls{MTE} (\Cref{sec:arm-mte}).
At run-time, all unsafe allocations (\dCTwo and \dCFour in \Cref{fig:stackframe}), and pointers based on them, are tagged as unsafe by clearing the topmost tag bit; resulting in a tag of form \unsafeTag.
We use the default tag, \safeTag, for safe allocations and unallocated memory that is initially tagged by the kernel.
The \guarded allocations are similarly tagged as safe, but are surrounded by differently tagged guards---either a dedicated  ``red zone'' or an unsafe allocation---to prevent unsafe dereferences.
All unsafe allocations and pointers are tagged with the unsafe tag, \unsafeTag, tag to isolate them from the memory-safe allocations.

If \theAttacker overwrites a pointer, it can be set to point to an arbitrary address, but also to have an arbitrary tag value, including the safe tag \safeTag.
To prevent the use of such pointers, we implement \emph{\tagProt} by clearing the topmost tag bit of a loaded pointer, unless it was loaded from a pointer-safe allocation (\Cref{sec:design-store-pointers}).
The address tag of an unsafe pointer could also be corrupted by large-enough indexing errors or pointer arithmetic that affects the high-order address bits of the pointer.
To prevent such pointer operations from changing address tags, our \tagProt stores the address tag before arithmetic operations, and then afterwards writes it back to the pointer.
Consequently, even if an unsafe pointer's address tag can be corrupted or a new pointer injected, the pointer will always be re-tagged as unsafe before use (\ref{req:forgery}).

\subsection{Realizing \guarded allocations}%
\label{sec:design-guarded}%
\label{sec:overflow-protection}

Buffer overflows are a common memory error that can be completely prevented with \gls{MTE}.
As long as the overflow is contiguous we can tag adjoining memory with a different tag to prevent the overflow (e.g., an overflow from \dCThree to \dCTwo in \Cref{fig:stackframe}).
For the purpose of our work, we consider an overflow linear if it is an under- or overflow such that its iteration step---e.g., how much an iterator is incremented on each loop---and element size guarantees that the overflow cannot ``hop'' over the adjoining memory granule and avoid detection.
We deem an allocation as \emph{\guarded} if it is not provably-safe, but can be shown to only allow linear overflows.
The adjoining memory can be either another allocation, e.g., a local variable, or a compiler-inserted guard zone.
We conservatively only assume that the adjoining 16-byte \gls{MTE} granule is tagged differently.
The stack layout can be optimized to avoid dedicated guards when stack variables can be reordered (\Cref{sec:memory-guards}).

\subsection{Detecting safely stored pointers}%
\label{sec:design-pointer-safe}%
\label{sec:design-store-pointers}

Implicit compiler-generated allocations are safe as they are only used by the compiler and not exposed to the programmer.
For instance, the return address (\dCOne in \Cref{fig:stackframe}) is stored in memory, but is not directly accessible by program code.
Another example are stack slots for local variables that are not explicitly referenced by the programmer (i.e., that do not ``have their address taken'') or ``register spills'' generated by the compiler to temporarily free registers.
Such implicitly-safe allocations do not require analysis, and will use the default safe tag.

In other cases, we must either assume that a pointer stored in memory is unsafe, or be able to exhaustively prove that its storage location is a safe allocation and that we can determine where it is subsequently loaded from memory.
However, when considering the safety of pointers in memory, memory-safety as defined above is insufficient.
We must also consider whether intra-object overflows could corrupt a pointer without violating the allocation boundaries of the storage location.
For instance, an array in a data structure can potentially overflow and corrupt pointers in the structure without violating the safety of the allocation as a whole\processifversion{NotAbridged}{ (\Cref{lst:allocation-bound-problem})}.
In addition to safety, we thus also consider the \emph{pointer-safety} of an allocation and deem an allocation to be \emph{pointer-safe} only if it is both safe and that pointers stored within it are safe from corruption.
Otherwise, we say the allocation is \emph{pointer-unsafe}, and assume that any pointers stored within it might be corrupted even if the allocation as a whole is safe.

\begin{NotAbridged}
  \begin{lstlisting}[language=c,label=lst:allocation-bound-problem,float=,floatplacement=tp,caption={%
    If a variable of type \inlinec{struct s} is memory safe, its use is guaranteed to never corrupt memory outside its bounds or after its lifetime.
    However, we must additionally verify that \inlinec{s.ptr} cannot be corrupted by an overflow of \inlinec{s.buff} without violating the memory safety of \texttt{struct s}.
    }]
  struct s { int *ptr; char buff[SIZE]; }
\end{lstlisting}
\end{NotAbridged}

\subsection{Compartmentalization}

All memory within the program is assigned to different safety types as shown in \Cref{tbl:design-summary}.
We assign tags to allocation classes so that we can isolate unsafe allocations, but also to facilitate \tagProt at runtime.
Specifically, whenever a pointer is loaded from a memory address, we can use the tag of the address to determine whether the loaded pointer is unsafe (shown in the~\ref{col:ptr-load} column of \Cref{tbl:design-summary}).
Any pointer loaded from memory not tagged as pointer-safe, is tagged as unsafe, and so can only access unsafe memory. 
Note that pointer-safe allocations may contain unsafe pointers, but because the location is pointer-safe, we do not need to perform \tagProt when loading those pointers.

\begin{table}[tp]
	\centering
	\begin{tabular}{|c|c|c|c|c|}
		\hline
		\multicolumn{2}{|c|}{Safety class/type} &
		 	\multicolumn{3}{c|}{Instrumentation}
		\\
		\newtag{Safe}{col:safe} &
			\newtag{Ptr. safe}{col:pointer-safe} &
			\newtag{\emph{tag}}{col:tag} &
			\newtag{\emph{guarded}}{col:guarded} &
			\newtag{\emph{ptr. load}}{col:ptr-load}
	  \\
		\hline\hline
		implicit & yes & \safeTag & --- & --- \\
		provable & yes & \safeTag & --- & --- \\
		provable & no & \ptrUnsafeTag & --- & unsafe \\
		\guarded & yes & \safeTag & guarded & --- \\
		\guarded & no & \ptrUnsafeTag & guarded & unsafe \\
		no & no & \unsafeTag & --- & unsafe \\
		\hline
	\end{tabular}%
	\caption{%
		The safety classes and corresponding run-time instrumentation.
		\ref*{col:tag} shows the tag set on allocation. 
		\ref*{col:guarded} indicates if guards are added.
		\ref*{col:ptr-load} shows if pointers loaed from the allocation are subject to\tagProt.
		\processifversion{NotAbridged}{%
		    Note that the~\ref*{col:tag} is statically assigned to the pointer at allocation and then allows us to determine the~\ref*{col:ptr-load} action at run-time.
		}
	}%
	\label{tbl:design-summary}
\end{table}

%% file: sections/060.analysis.tex
\section{Memory Safety Analysis}%
\label{sec:analysis}

We implement our memory safety analysis as two LLVM \gls{IR} passes extending the functionality of the existing stack safety analysis (\Cref{sec:llvm-ssa}).
It analyzes the use of stack allocations represented as \inlinellvm{alloca} instructions in the \gls{IR}.
Recall that implicitly-safe allocations---i.e., those that are not represented by \inlinellvm{alloca} instructions---are always safe and are not analyzed.
For all other stack allocations, the analysis indicates whether it can prove the allocation to be either safe (or \guarded), or whether it must be assumed to be unsafe.
The analysis assumes instrumentation that enforces isolation (\Cref{sec:design-fault-isolation}) and prevents linear overflows (\Cref{sec:overflow-protection}).
Given such instrumentation, we can use an analysis to find allocations that:
\begin{enumerate}
  \item are memory safe;
  \item can be \guarded such that their safety can be guaranteed at run-time; and,
  \item are pointer-safe such that pointers within them are safe, and the safety of all pointers loaded from it can be verified.
\end{enumerate}
Programmatic analysis of any non-trivial program properties is undecidable~\cite{riceClasses1953}.
Consequently, most static analyses, including ours, are approximate.
For security, the analysis must be conservative but can over-approximate the unsafe set of allocations.
Our goal is to not fully analyze all memory use, but to maximize the set of allocations that can be proven memory safe (\Cref{sec:memory-safety-def}).
In this work, we present an example analysis building upon the stack safety analysis by~\cite{kuznetsovCodepointer2014}, but our instrumentation strategy is not tied to this specific analysis.

\begin{algorithm}
\begin{algorithmic}
  \State{} \( UseInfo \gets \texttt{new UseInfo} (base\_pointer) \)
  \State{} \( WorkList \gets base\_pointer \)

  \While{$ !WorkList.\Call{empty} $}
    \State{}$ Ptr \gets WorkList.\Call{pop} $
    \ForAll{$ Use \in Ptr.\Call{uses} $}
      \If{$ Use.\Call{definesNewPointer} $}
        \State{} \( WorkList \gets Use \) \Comment{add ptr to list}
      \ElsIf{ \( Use \in CallInst \)}
        \State{} \( UseInfo.calls \gets Ptr \)
      \ElsIf{ \( Use \in LoadInst \)}
        \State{}$ UseInfo.range \gets \Call{getRange}{Use} $
        \State{}$ UseInfo.DerefedBy \gets \Call{getLoadInfo}{Use} $
      \ElsIf{ \( Use \in StoreInst \)}
        \If{$ Use.\Call{isPointerOp}{Ptr} $} \Comment{ptr used}
          \State{}$ UseInfo.range \gets \Call{getRange}{Use} $
        \Else{} \Comment{otherwise, ptr itself is stored}
          \State{} \( UseInfo.StoredIn \gets PtrOp \)
        \EndIf{}
			\Else{}
      	\State{} \( UseInfo.\Call{setToUnsafe}{} \)
      \EndIf{}
    \EndFor{}
  \EndWhile{}
\end{algorithmic}
\caption{%
For each analyzed pointer, the \inlinec{FunctionPass} creates a \UseInfo.
It then uses the def-use chain to find all uses of the allocation and update \UseInfo accordingly.
}%
\label{alg:analysis-basic}
\end{algorithm}

Our analysis consists of a local intra-procedural \inlinec{FunctionPass} and a later inter-procedural \inlinec{ModulePass} that collects and merges the local results.
The goal is to analyze allocations, the \inlinec{FunctionPass} tracks three types of pointers: \begin{enumerate*} 
	\item pointers to local allocations,
	\item function arguments that are pointers, and
	\item pointers loaded from memory.
\end{enumerate*}
We call these such pointers \emph{base pointers}.
For each \emph{base pointer} our analysis follows the use of that pointer and stores the information in a \UseInfo container object.
We reuse the \UseInfo from \StackSafetyAnalysis but extend it with information needed to support our analysis.
When a pointer is derived from the base pointer of a \UseInfo, we refer to it as being \emph{based on} the \UseInfo.

The \inlinec{FunctionPass} does not track pointers passed to called functions or stored in memory, instead it only records such events in the corresponding \UseInfo.
For each \UseInfo \texttt{U}, the later \inlinec{ModulePass} then either finds and merges the corresponding \UseInfo for the calls, stores, and loads; or marks \texttt{U} as unsafe.
We stop processing a \UseInfo if it is shown to be unsafe.
The \inlinec{ModulePass} is iterative and continues until no further updates take place.
To handle cyclic function calls and to limit analysis complexity, the analysis depth is limited such that any \UseInfo reaching the limit is marked as unsafe.

\subsection{Detecting safe allocations}%
\label{sec:analysis-fault-isolation}%
\label{sec:analysis-structure}%
\label{sec:analysis-safe}%
\label{sec:analysis-casting}

\begin{NotAbridged}
The local analysis starts by generating a \UseInfo for \begin{enumerate*}
	\item all pointers to local allocations (defined by \inlinellvm{alloca} instructions),
\item all pointer arguments to the function (as defined by the function definition), and
\item any pointer loaded from memory (any \inlinellvm{load} instruction).
\end{enumerate*}
\end{NotAbridged}
The \inlinec{UseInfo} is used to store information on all uses of the base pointer or pointers based on it.
As the LLVM \gls{IR} is structured in \gls{SSA} form, we can use the base pointer's definition-use chain---which links variables to any instruction that uses them---to track its use throughout the function, as shown in  \Cref{alg:analysis-basic}.
When a pointer is used to access memory, the range of that access is added to the \inlinec{UseInfo}.
When a pointer is passed to another function, the called function and offset to the base pointer is recorded in the \inlinec{UseInfo}.
When a new pointer is derived (for instance when incrementing an iterator), the new pointer is stored for subsequent analysis.
Finally, to support tracking pointers in memory, we also record any pointer loads in the \inlinec{UseInfo.DerefedBy} list and pointer stores in the \inlinec{UseInfo.StoredIn} list.
If the analysis cannot handle some use of a pointer---for instance, a memory access with an unknown offset---the corresponding \UseInfo is marked as unsafe and further analysis is stopped.
As the \inlinec{FunctionPass} is local, it cannot resolve how pointers passed to called functions are used and instead only stores information about the call for later analysis.
We also defer resolving the use of pointers stored in memory to the \inlinec{ModulePass}.

\Cref{lst:analysis-example} shows an example of two functions that can only be partially resolved locally.
During the \inlinec{FunctionPass}, only \inlinec{A} in \inlinec{func\_1} can be determined to be safe, as no references to it are passed outside the local scope or written to memory.
In contrast, a reference to \inlinec{B} is stored in memory and then passed to the called function, so its safety cannot be determined before the \inlinec{ModulePass}.
When analyzing \inlinec{func\_2}, the \inlinec{ptr} argument can be fully analyzed because it is not written to memory or used outside the local scope.

\begin{lstlisting}[language=c,float=,floatplacement=tp,caption={
  \processifversion{Abridged}{%
    Function with variables of different security classes.
  }
  \processifversion{NotAbridged}{%
    A function containing three local variables \inlinec{A}, \inlinec{B}, and the pointer \inlinec{ptr}.
    The safety of \inlinec{A} can be locally determined.
    However, a pointer to \inlinec{B} is passed via \inlinec{ptr} into \inlinec{func\_2()}; its safety depends on the safety of the corresponding argument of \inlinec{func\_2()} and how it is used when loaded from.
  }
},label=lst:analysis-example,linerange={5-16}]
void func_1(int **arg) {
    int A, B;
    int *ptr;
    A = 0;        // Safe assignment to A
    ptr = &B;     // Create pointer to B
    func_2(&ptr); // Pass pointer to ptr
    func_2(arg);
}
void func_2(int **ptr) {
    for (int i = 0; i  < 10; ++i)
        (*ptr)[i] = 0; /* \label{line:analysis-example:loadinfo} */
}
\end{lstlisting}

The \inlinec{ModulePass} resolves any calls or stores of potentially safe allocations until each allocation is either exhaustively analyzed and proven safe, or assumed to be unsafe.
It works by adding all functions to a work list and then processing the list one function at a time, as shown in~\Cref{alg:analysis-global-pass}.
Within a function, every \UseInfo \texttt{U} is then processed.
For each function call in \texttt{U.calls}, we query the analysis results for the corresponding argument \UseInfo in the called function and merge it to \texttt{U}.
For each memory store in \texttt{U.StoredIn}, we query the \texttt{UseInfo.DerefedBy} list (populated by~\Cref{alg:analysis-basic}) of that storage location and merge these to \texttt{U}.
If a pointer is stored or loaded in an unsafe location, or if we cannot find \UseInfo for all possible store or load locations of \texttt{U}, then \texttt{U} is marked unsafe.

When merging a \UseInfo to \inlinec{U}, we add the memory access ranges, the \texttt{calls}, \texttt{StoredIn}, and \inlinec{DerefedBy} lists to \inlinec{U}.
If the processing of a function causes changes in any of its \UseInfo then all of its callers are re-inserted on the top of the work list.
The \inlinec{ModulePass} only works with \UseInfo objects and does not need to process the LLVM \gls{IR} again.

\begin{algorithm}
\begin{algorithmic}

  \Function{runModulePass}{$ CallersMap, WorkList $}
    \While{$ !WorkList.\Call{isEmpty} $}
      \State{}$ F \gets WorkList.\Call{pop} $
      \If{$ F.!\Call{incIfLessThan}{LIMIT} $}
        \State{}$ F.\Call{setAllToUnsafe}{U} $
      \Else{}
        \ForAll{$\texttt{UseInfo } U \in F$}
          \State{}$ OriginalU \gets U $

          \ForAll{$ C \in UseInfo.Calls $}
            \State{}$ U.\Call{mergeCalleeUseInfo}{C} $
          \EndFor{}

          \ForAll{$ C \in UseInfo.StoredIn $}
            \ForAll{$ L \in C.derefedBy $}
              \State{}$ U.\Call{mergeLoadUseInfo}{L} $
            \EndFor{}
          \EndFor{}

          \If{$ (OriginalU \ne U) $}
            \State{}$ WorkList.\Call{pushAll}{Callers[F]} $
          \EndIf{}
        \EndFor{}
      \EndIf{}
    \EndWhile{}
  \EndFunction{}

\end{algorithmic}
\caption{%
The \inlinec{ModulePass} processes functions one at a time by updating their contained \inlinec{UseInfo} objects.
}%
\label{alg:analysis-global-pass}
\end{algorithm}

For example, in \Cref{lst:analysis-example}, the \inlinec{ModulePass} will update \inlinec{func\_1} by going through its set of \inlinec{UseInfo}.
As \inlinec{ptr} has \inlinec{func\_2} in its calls set, the \UseInfo of the corresponding argument of \inlinec{func\_2} is merged to it.
The merge includes not only the immediate access range of \inlinec{ptr}, but also the additional information pertaining to function calls and memory stores/loads; in this case, from the \UseInfo for the pointer dereference at \Cref{line:analysis-example:loadinfo}.
Consequently, subsequent updates will further propagate this update to the \UseInfo of \inlinec{B} as well.

During the analysis, the pointed-to memory location of a pointer is typically unknown.
In particular, during the \inlinec{FunctionPass}, pointer arguments passed to the function remain unknown and may point to different memory depending on execution / caller context.
To accommodate this, all ranges are treated as \emph{offsets} to the value of the base pointer of the \UseInfo.
When merging two \inlinec{UseInfos} we then apply the offset to the added range before applying it to the destination \UseInfo.
For instance, if a pointer to the 10th element of an array is passed to a function, then the resulting range is offset by 10 when merging the \UseInfo of the argument to the caller's corresponding \UseInfo.
At any point, if an offset or range cannot be statically determined, the \UseInfo is marked as unsafe.
Consequently, at the end of analysis, the \UseInfo of each stack allocation either ends up being  marked as unsafe, or alternatively it can be used to retrieve the memory access range to verify that any use of a pointer based on the allocation remains within the bounds of the allocation size.

\subsection{Detecting linear-overflows}%
\label{sec:analysis-guarded}

\begin{figure}
\centering
\includegraphics[width=0.8\columnwidth]{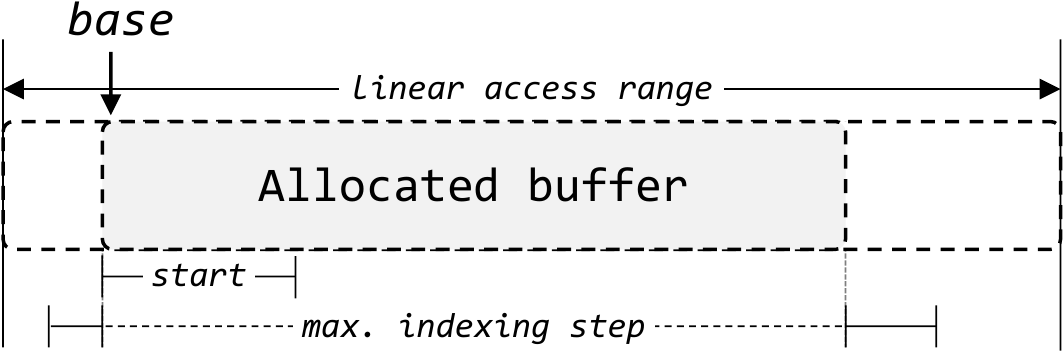}
\caption{
  The static safety analysis is extended to also track the linear access range and maximum index step.
  \processifversion{NotAbridged}{%
    These allow us to determine when overflow protection is needed, and when an overflow is guaranteed to hit our guard page.
  }
}\label{fig:linearprotection}
\end{figure}

In addition to provably memory-safe allocations, our analysis also recognizes \guarded allocations that can be instrumented to be effectively memory safe or crash at run-time (\Cref{sec:design}).
For this, we extend \UseInfo to separately track the \emph{arbitrary access ranges} and the \emph{linear access range} (\Cref{fig:linearprotection}).
To determine whether linear overflows can be prevented using \gls{MTE}, we also store the information necessary to determine
\begin{enumerate*}
  \item whether the linear access starts within bounds, and
  \item whether the increment / decrement size is less than the used \gls{MTE} memory-tag granule size used by the instrumentation.
\end{enumerate*}
For example, in \Cref{lst:code-example}, \inlinec{buf\_lin} is always first accessed within bounds, and guaranteed to hit a different \gls{MTE} memory tag because a single loop iteration increments the address by \inlinec{sizeof(unsigned)} bytes which is smaller than the \gls{MTE} tagging granule of 16 bytes.

The memory access range analysis uses the LLVM's scalar evolution framework to determine linear access behavior.
For instance, in a loop over an array, the scalar-evolution indicates the first element accessed and how the index changes---or, evolves---on each iteration of the loop.
If the start and indexing steps are known and bounded, we treat the access as linear and update the linear access range, start range, and indexing step in \UseInfo.
Otherwise, we treat the access as non-linear and only update the arbitrary access range. 
The linear access information is updated along with other ranges, as specified in \Cref{alg:analysis-basic,alg:analysis-global-pass}.
Consequently, as long as the arbitrary access range and linear access start range of a \UseInfo remain within the bounds of the allocation, and the indexing step remains bounded, we can designate an allocation to be \guarded, even if it would not otherwise be memory safe.

\subsection{Tracking pointers stored in memory}%
\label{sec:analysis-pointer-safe}

Our safety analysis can also reason about pointers stored in memory when the strong isolation provided by \gls{MTE} can guarantee that such pointers remain uncorrupted.
Where in the general case we need a full program analysis to reason about memory safety, the isolation guarantees allow for a more lightweight local analysis. 
As shown in \Cref{alg:analysis-basic}, when our \inlinec{FunctionPass} encounters a pointer is stored in memory, we mark it as unsafe unless we show that the storage location is safe.
Specifically, we look at the \inlinellvm{StoreInst} and use the use-def chain to verify that all possible pointer operands of the instruction have a corresponding UseInfo we can prove as safe.
We mark the \UseInfo unsafe if the use-def traversal encounters non-local memory; for instance, if the pointer operand is a global variable.
\begin{Abridged}
  To detect inter-object overflows, we further mark a \UseInfo as unsafe if it is passed to an incompatible type or any memory accesses using it violate the type boundaries of the allocated type.
\end{Abridged}
Consequently, either each possible storage location for a pointer based on an analyzed \UseInfo \texttt{U} has a corresponding \UseInfo in the \inlinec{U.StoredIn} set, or otherwise, \texttt{U} is marked unsafe.
By constraining this analysis to local scope, we efficiently reach sound results without the need for complex full-program safety analysis.

\begin{NotAbridged}
For pointers stored \emph{within compound} structures the measures listed earlier do not provide safety guarantees.
As shown in \Cref{lst:allocation-bound-problem}, such pointers might be corruptible without violating the allocations bounds.
We address this problem by approximating pointer-safety as typed use.
The analysis we have implemented so far uses a less precise (but still conservative) approximation that assumes that only allocations of pointer-type are pointer-safe.
However, we plan to extend our implementation based on the following algorithm that utilizes type information:
We assume that an allocation is \emph{pointer-safe} if we can determine that pointer fields in a compound structure are always used in a typed manner and that any other use of the structure cannot overflow to those pointer fields.
For example, considering an allocation of type \inlinec{struct s} in \Cref{lst:allocation-bound-problem}, our algorithm verifies that all pointers based on the base pointer retain the type of the pointer constituents of the structure.
For instance, if a pointer to \inlinec{struct s} is cast to a plain \inlinec{char *}, we lose the type of the contained pointer \inlinec{s.ptr} and mark the \inlinec{UseInfo} it is based on as \emph{pointer unsafe}.
For each memory access, we can then verify that either it is a typed use of the pointer, or if not, whether the merged access range is guaranteed to not modify pointers within the structure.
For instance, if \inlinec{s.buff} is used, and we cannot prove it will not overflow, we mark the \inlinec{UseInfo} it is based on as pointer unsafe.
In both cases, pointer-safety always depends on the allocation being memory safe to begin with.
This conservative approach allows us to locally over-approximate pointer-safety violations during our \inlinec{FunctionPass}, and then, in the \inlinec{ModulePass}, propagate the pointer-safety state along with other \UseInfo state.
\end{NotAbridged}

\subsection{Analysis results and use}

Our instrumentation uses the analysis results to assign each stack allocation defined by an \texttt{alloca} to one of the safety and pointer-safety types shown in \Cref{tbl:design-summary}.
\begin{NotAbridged}
All implicitly safe allocations will be materialized when the compiler lowers the \gls{IR} into Machine \gls{IR}; and as mentioned, need not be further analyzed.
\end{NotAbridged}
Our analysis results only apply to allocations, and do not pertain to the run-time safety of specific pointers; which might, at run-time, point to both safe and unsafe pointers at different times.
Instead, the instrumentation utilizes the analysis to assign tags at allocation-time such that the \tagProt can ensure that only pointers legitimately based on safe allocations can be dereferenced to safe allocations.

%% file: sections/070.implementation.tex
\section{Instrumentation}%
\label{sec:implementation}

\begin{figure}[tp]
    \centering
    \includegraphics[width=1\columnwidth]{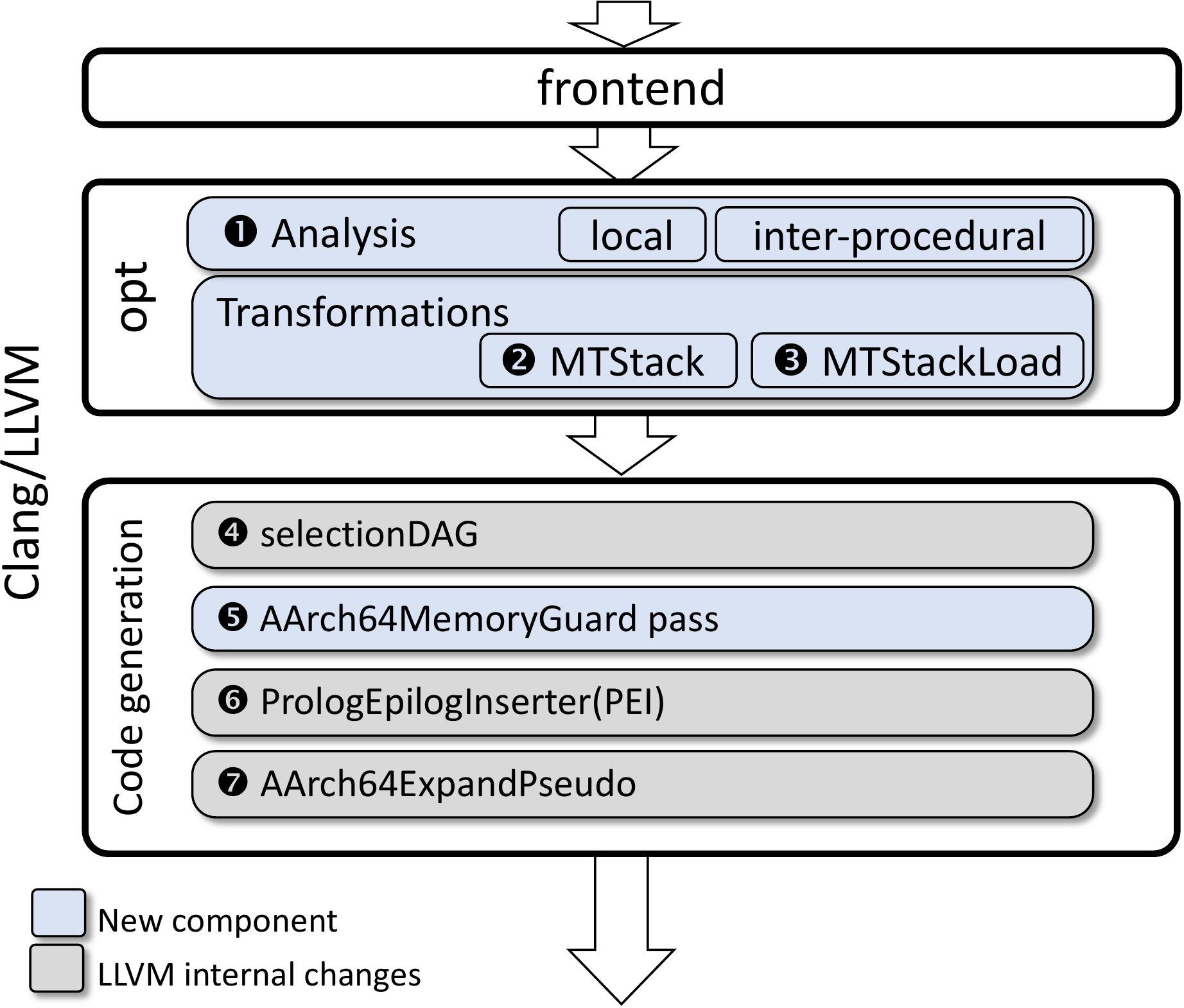}
    \caption{%
        \processifversion{Abridged}{LLVM opt and AArch64 backend changes.}
        \processifversion{NotAbridged}{%
		    The analysis and instrumentation adds passes and functionality to both the \gls{IR} optimizer and the architecture-specific backend.
		    The stack safety analysis is replaced by our \gls{MTE}-aware stack safety analysis (\Cref{sec:analysis}), and the instrumentation is split into a number of modified and new passes outlined in \Cref{sec:implementation}.
		}%
		}%
    \label{fig:compiler-mte}
\end{figure}

Our instrumentation is based on \LLVMversion.
\Cref{fig:compiler-mte} shows an overview of the compiler pipeline and stages modified by our work: \begin{enumerate*}
\item the analysis passes described in \Cref{sec:analysis} (\dOne),
\item two new transformation passes (\dTwo, \dThree), and
\item changes to machine code generation (\dFour--\dSeven).
\end{enumerate*}
We also modify the libc runtime library to enable \gls{MTE} at the start of the process.

\subsection{Protecting safe allocations}%
\label{sec:impl-fault-isolation}

Our new \gls{IR} pass, \MTStack, is responsible for tagging both pointers and memory allocations (\Cref{sec:MTStackOpt}). 
It runs our memory safety analysis and uses the analysis results to guide which allocations to tag (\Cref{sec:design-fault-isolation,sec:analysis-fault-isolation}).
We use the default tag for safe allocations.
Consequently, \MTStack only needs to tag unsafe allocations and pointers based on them, whereas the kernel will initially tag the stack with the default tag.
On mainline Linux, the default tag is set to $\mathtt{0b0000}$~\cite{frascinoMemory2020}.
We change the default tag to \safeTag, which avoids the need for special handling of NULL pointers that would otherwise unintentionally have the safe tag.
A second \gls{IR} pass, \MTStackLoad, then instruments unsafe pointer operations and loads to realize \tagProt (\Cref{sec:impl-tag-prot}).

\subsubsection{Tagging}%
\label{sec:MTStackOpt}

The \MTStack pass initially instruments the \gls{IR} \inlinellvm{alloca} calls that generate stack allocations.
For each stack allocation \MTStack checks if it is unsafe, and if so, inserts a \inlinellvm{llvm.aarch64.settag} intrinsic that tags the allocation.
Compiler intrinsics are compiler-defined functions that are lowered to architecture-specific code during machine code generation.
Consequently, the compiler can optimize intrinsics based on their defined high-level semantics while deferring the detailed machine code generation to later.
We minimize register pressure when lowering the \gls{IR} by marking the \instruction{addg} and \instruction{subg} instructions that generate tagged pointers as re-materializable; i.e., such that tagged pointers can be re-created instead of needing to be stored.

To prevent temporal memory errors, \MTStack resets the tags of all unsafe allocations when the stack frame is released on function return.
\begin{NotAbridged}
This prevents the use of dangling pointers to dead stack frames.
For instance, in \Cref{lst:code-example}, a pointer to the stack allocation on \Cref{line:code-example:alloca-to-global} is stored in the global variable \inlinec{g\_ptr}, and could be used after the function returns.
\end{NotAbridged}
Moreover, resetting tags also ensures that unallocated memory is returned to its default state.
\begin{NotAbridged}
Recall that safe allocations use the default tag.
Therefore, those need not be explicitly tagged by \MTStack.
\end{NotAbridged}

\begin{NotAbridged}
  In most cases, the stack frame size is fixed , and hence our tagging strategy can be static.
  However, \MTStack must also tag \glspl{VLA} and memory allocated using the C \inlinec{alloca} function.
  In these cases, \MTStack uses the same \inlinellvm{settag} intrinsic, but our modified instruction selection \inlinec{SelectionDAG} lowers the \inlinec{settag} to a new pseudo-instruction \inlinellvm{STGLoopGeneric} when either the allocation size is unknown or multiple 16-byte granule tag writes are needed.
  Eventually, the \inlinec{AArch64ExpandPseudo} then expands the pseudo-instruction to the machine code shown in \Cref{lst:stgloopgen}.
  This code is currently inlined, but to minimize code size, we plan to add a generic function to the C runtime for this purpose.

\begin{lstlisting}[language=aarch64,float=,floatplacement=tp,
  caption={%
	The expanded AArch64 machine code for the \inlinec{STGLoopGeneric} pseudo-instruction used to dynamically tag memory ranges.
	We use both the \instruction{STG} and \instruction{ST2G} instructions to minimize the number of separate tag memory writes.
	},
 label={lst:stgloopgen}]
    tbz     x10, #4, .sizecheck
    stg     x9, [x9], #16   ; range start
    sub     x10, x10, #16   ; range size
.sizecheck:
    cbz     x10, .done
.loop:
    st2g    x9, [x9], #32
    sub     x10, x10, #32
    cbnz    x10, .loop
\end{lstlisting}
\end{NotAbridged}

The C \inlinec{alloca} poses a challenge because the allocated memory can be dynamically sized and is freed only at function exit.
An \inlinec{alloca} can be anywhere---including in conditional blocks---and it may not always be executed before return.
Consequently, we do not always statically know whether, and how much, memory is allocated on the stack.
\MTStack marks such functions with a new \inlinellvm{reset-tags} attribute, that is used by the \gls{PEI} to inject an intrinsic that re-tags the whole stack frame before return.

\begin{NotAbridged}
  Our \gls{VLA} and C \inlinec{alloca} instrumentation also prevents attacks where such allocations overflow the whole stack in order to access the heap.
  Such overflows are commonly mitigated by following the stack memory with an unmapped memory page that causes a fault if written to.
  Because this only happens on write, not allocation, an attacker can avoid this by not writing to the unmapped page.
  In contrast, our instrumentation prevents this attack altogether as a fault would be triggered when the memory is initially tagged.
\end{NotAbridged}

Finally, recall, that on ARMv8.5 \gls{MTE} the granule size is 16 bytes (\Cref{sec:background-mte}).
Therefore, the instrumentation must pad and align all tagged allocations to the granule size.
For alignment, \MTStack uses the existing \inlinellvm{align} attribute to set the alignment of allocations in the \gls{IR}.
To avoid unnecessary transformation of the \gls{IR}, we do not add padding in \MTStack.
Instead, allocations that must be padded are marked with our new \inlinellvm{tagged} attribute.
We then extend the \inlinec{MachineFrameInfo} to resize the marked stack slots based on the \inlinellvm{tagged} attribute.

\subsubsection{\TagProt}%
\label{sec:impl-tag-prot}%
\label{sec:impl-tag-forgery-protection}%
\label{sec:impl-casting}%
\label{sec:null-pointer-instrumentaiton}

\begin{NotAbridged}
As described in \Cref{sec:design-tag-prot}, we must prevent an attacker exploiting unsafe allocations, pointer casts, or pointer arithmetic to forge tagged pointers to safe memory allocations.
\end{NotAbridged}
\MTStackLoad ensures that all corrupted pointers are tagged as unsafe before their use, and that pointer arithmetic cannot be used to modify address tags.
If the pointer was loaded from unsafe memory, \MTStackLoad clears the topmost bit in the address tag of the pointer before it is dereferenced.
As a result, either the loaded pointer was stored in safe memory and its address tag remains unchanged; or it was loaded from unsafe memory and the topmost tag bit will be unset before it is used to access memory.

A common pattern is to use  pointer ``sentinel'' values---e.g., NULL---to check an object is allocated before use.
\processifversion{Abridged}{Since \tagProt interferes with this, our instrumentation retains the unmodified pointer for use in comparisons.}
\begin{NotAbridged}
  To support such cases, we use the unmodified pointer for comparisons, and 
  While the NULL pointer is unaffected by \tagProt, we have also observed other sentinel values such as \texttt{-1} that will be changed when the tag bit is unset.
  We deal with this corner case by ensuring that comparison operations use the original unmodified pointer to detect possible sentinel values.
  To do so, we use a local-data flow analysis on loaded unsafe pointers, and selectively substitute the pointer with the enforced variant before it is used to dereference memory or used as a function argument.
  We do not track sentinel values globally and instead apply \tagProt when they are passed as function parameters by value.
  This is a rare corner case and we left its implementation for future work as no such instance was encountered in any tested code.
  If an unsafe pointer is stored in memory, we know its new storage location is going to be pointer unsafe, and can safely store the original unmodified pointer to ensure it is correctly handled when subsequently loaded again.
\end{NotAbridged}

\begin{NotAbridged}
As future work, \MTStackLoad could optimize \tagProt use by omitting instrumentation where safety can be statically determined.
When an allocation is known to be unsafe, the topmost tag bit can be unconditionally set on any pointer loaded from it.
Conversely, when an allocation is known to be pointer-safe, we need not instrument loads from it.
Such optimizations cannot be done if the safety of the pointer depends on execution context.
\end{NotAbridged}

\MTStackLoad prevents tag forgery via (overflowing) pointer arithmetic.
To do so, \MTStackLoad inspects \gls{GEP} instructions that are used to represent pointer arithmetic in the \gls{IR}.
Either the \gls{GEP} can be statically determined as correct, or it is instrumented to prevent it from changing address tags.

\begin{NotAbridged}
A special case that avoids the \gls{GEP} \tagProt is casting a pointer to an integer before manipulating it, and subsequently casting it back to a pointer.
Such casting of integer-typed pointers are handled by treating any pointer defined by a cast from non-pointer type as unsafe, and unconditionally setting the topmost tag bit to zero after the cast.
Correspondingly, whenever the analysis encounters a pointer being cast to an integer, it will mark the allocation it is based on as unsafe.
\end{NotAbridged}

\subsection{Preventing linear overflows}%
\label{sec:impl-guarded}%
\label{sec:memory-guards}

To realize the \guarded safety type, \MTStack inserts memory guards around \guarded allocations (\Cref{sec:design-guarded,sec:analysis-guarded}).
The tags of a memory guard and its guarded allocation are different.
The guard can be either additional unused memory or an existing allocation with a different tag (see Figure~\ref{fig:stackframe}). 

We insert memory-guard allocations and mark them as such using our \inlinellvm{llvm.aarch64.memoryguard} intrinsic that prevents other passes from removing them because they appear unused.
During machine code generation, the intrinsic is used to identify the stack slot and to mark it as a memory guard through an added interface in the \inlinec{MachineFrameInfo} class.
Later, the \inlinec{AArch64MemoryGuard} pass will remove redundant memory guards, for example, if there is already a differently tagged unsafe allocations adjoining a \guarded allocation, then an explicit guard is not needed between them.
\begin{NotAbridged}
As future work we plan to further optimize stack memory consumption by reordering allocations to minimize the need for explicit memory guards.
\end{NotAbridged}

\subsection{Pointer-safe tagging}%
\label{sec:imple-pointer-safe}

Recall that safe allocations could still allow inter-object corruption unless it is also pointer-safe (\Cref{sec:design-pointer-safe,sec:analysis-pointer-safe}).
To distinguish such safe, but pointer-unsafe allocations, we tag them using the \ptrUnsafeTag.
Consequently, we can at run-time distinguish pointers loaded from pointer-safe allocations, and apply \tagProt to all other loaded pointers.

%% file: sections/080.evaluation.tex
\section{Evaluation}%
\label{sec:evaluation}

We evaluate our design and implementation described in \Cref{sec:design,sec:analysis,sec:implementation} in terms of security, functionality, and performance with respect to the requirements defined in \Cref{sec:requirements}.

\subsection{Security evaluation}%
\label{sec:security-evaluation}

To show that safe allocations cannot be corrupted (\ref{req:safe}), we show that pointers based on unsafe allocations cannot corrupt them (\ref{req:compart}) and that \theAttacker cannot forge safe address tags (\ref{req:forgery}).

Our memory safety analysis is conservative (\Cref{sec:analysis}), but not necessarily complete.
Hence, we might not find \emph{all} safe and pointer-safe allocations, but ensure that those found are, in fact, safe.
To do so, an allocation and associated pointers are marked as safe only if we can track all references to it, and verify that those pointers are safely used (\Cref{alg:analysis-basic,alg:analysis-global-pass}).
Consequently, we know that a pointer is tagged as safe only if it based on a safe allocation and cannot corrupt or overflow other allocations (\Cref{sec:design}).
Conversely, we know that any other pointer is initially tagged as unsafe and prevented from accessing safe allocations by the \gls{MTE} hardware.
To ensure~\ref{req:compart} we must also prevent \theAttacker from forging or altering pointers such that they have the safe tag.

The \tagProt feature of \IMPLNAME{} prevents \theAttacker from introducing new pointers with safe tags via pointer corruption or injection (\Cref{sec:design-tag-prot,sec:impl-tag-forgery-protection}).
Specifically, we ensure that:
\begin{enumerate*}
	\item the first tag-bit is cleared when casting a non-pointer to a pointer (thus tagging it as unsafe),
	\item unsafe pointer arithmetic cannot change the address tag, and
	\item any pointer loaded from pointer-unsafe memory is tagged as unsafe.
\end{enumerate*}
Consequently, while \theAttacker can set pointer-unsafe pointers to arbitrary values, these are subject to \tagProt before they can be used.
To avoid \tagProt, \theAttacker could use intra-allocation corruption and inject forged pointers into pointer-safe memory without violating the allocation boundaries.
To address this, our analysis verifies that non-pointer writes cannot corrupt pointer fields in pointer-safe allocations (\Cref{sec:design-store-pointers,sec:analysis-pointer-safe}).
\theAttacker might still inject pointers as values into non-pointer fields; but in this case, the value will be subject to \tagProt when it is cast to a pointer before dereference.
Consequently, \tagProt stops \theAttacker from accessing safe allocations using forged address tags, irrespective of whether those tags are directly manipulated or injected into memory (\ref{req:forgery}). 

\begin{NotAbridged}
Signal-handlers can break the typical control-flow of program.
However, they will be analyzed and instrumented similar to all other functions.
Consequently, all pointers used by them are local to the control-flow starting from the handler entry-point; or they are pointers to unsafe allocations passed to the handler via globally-accessible variables.
\end{NotAbridged}

We use alternating tag values to prevent buffer overflows.
Specifically, we tag adjacent allocations to realize the \guarded subset of the safe domain (\Cref{sec:impl-guarded}).
Our analysis verifies that any such \guarded overflow cannot ``jump'' past the alternating tags by incrementing the pointer more than the 16-byte memory granule (\Cref{sec:analysis-guarded}).
Consequently, such allocations can be safely added to the safe domain (\ref{req:overflow}).

We do not rely on random or hidden tags, hence \IMPLNAME{} is not affected by memory disclosures that leak memory tags (\ref{req:disclosure}).
Therefore, we conclude that our design meets requirements \ref{req:safe}--\ref{req:disclosure}.

\subsection{Functional verification}%
\label{sec:functional-evaluation}

For functional verification, we use \qemuVersion, which supports \gls{MTE}~\cite{qemu}.
The emulator replicates \gls{MTE} functionality on architectural level, and so, can be used to test compatibility between our instrumentation, \gls{MTE}, and existing source code.
We compiled fully instrumented variants of the evaluated SPEC CPU 2017 benchmarks~\cite{spec-cpu-2017} and executed them on \qemuVersion.
The functional verification consists of \emph{all} the C language benchmarks%
, which we also used for our performance evaluation (see Table~\ref{tbl:spec-overhead}).
Since the SPEC CPU 2017 benchmark suite consists of real-world programs, the evaluated tests are a strong indication that our instrumentation strategy is widely applicable (\ref{req:comp}).

\subsection{Performance evaluation}%
\label{sec:performance-evaluation}

Our performance evaluation covers three aspects of \IMPLNAME{} overhead:
\begin{enumerate*}
	\item the increase in \emph{code size} due to added instrumentation code, 
	\item the increase in \emph{stack use} due to the need to align and pad memory, and 
	\item the increase in \emph{execution time} due to changes by instrumentation and checks by the \gls{MTE} hardware.
\end{enumerate*}
We used the SPEC CPU 2017~\cite{spec-cpu-2017} C benchmarks shown in \Cref{tbl:spec-overhead} for performance evaluation.

We estimate code and stack size as follows:
The code size is estimated by comparing the \texttt{.text} section size of non-instrumented and fully instrumented benchmark binaries.
As the stack grows and shrinks during execution, maximum run-time stack use is not necessarily a useful metric.
Instead, we gather compile-time statistics and measure the average function stack-frame increase throughout the whole program.
Both the stack-frame size and \texttt{.text} size is determined based on fully \gls{MTE}-instrumented binaries.

The execution-time overhead cannot be directly measured without \gls{MTE} hardware or cycle-accurate emulators.
Therefore, we instrument the benchmarks with emulated \gls{MTE} overhead analogues as described below (\Cref{sec:mte-overhead-analogue}) and run them on a TaiShan 2280 Balanced Model~\cite{huaweiTaiShan2020} with a \benchmarkHardware ~\cite{xiaKunpeng2021}, an ARMv8 class-A CPU without \gls{MTE} support.
We measured the run time of each benchmark using the GNU/Linux \texttt{time} utility over 10 runs each.

\subsubsection{Emulating \gls{MTE}-instruction overhead}%
\label{sec:mte-overhead-analogue}

The performance overhead in terms of execution time consist of four components:
\begin{enumerate*}
	\item changes in code or memory layout,
	\item address tagging,
	\item memory tagging, and
	\item tag checks on memory access.
\end{enumerate*}
To measure the instrumentation overhead, we ensure that each component is measurable as-is, or by using an \emph{\gls{MTE} overhead analogue} to estimate an upper bound.
Our analogue is similar to the independently developed, MTE instruction analogue used by HAKC~\cite{mckeePreventing2022} in that we manipulate the tag bits of pointers to simulate address tags and replace memory tag writes with regular writes to memory.
We now describe in detail our \gls{MTE} overhead analogue used for each of these.%

\textbf{Code and memory layout} changes are caused by the need to align allocations to the \gls{MTE} granule of 16-bytes, increased register pressure, and other changes needed to accommodate the added code.
Where possible, we convert each \gls{MTE} instruction to an equivalent generic instruction with the same arguments.
Doing so minimizes code layout changes caused by the overhead emulation. 
In some cases, such as with \inlineasm{st2g}, we need multiple general-purpose instructions to emulate a single \gls{MTE} instruction.
However, this does not invalidate our upper bound as our code layout changes always increase code size, never decrease it.

\textbf{Address tagging} is emulated using regular bit manipulations to modify the address bit of a pointer.
The specialized \gls{MTE} instructions are expected to be at least as fast as the corresponding bit-manipulation realized with general purpose instructions.
Similar to any \gls{MTE} instrumentation, we then enable the \gls{TBI} feature (\Cref{sec:background-mte}) to ignore tag bits during address translation.

\textbf{Memory tagging} overhead is challenging to estimate without access to actual tag-memory implementations.
However, as these operations are specialized to modify tag memory (\Cref{sec:background-mte}) we can assume that they are optimized to take advantage of locality and the CPU caches.
Therefore, we use regular stores to represent an upper bound on the cost of writing to the specialized tag memory.
Specifically, we convert all memory tag writes to regular memory writes.
\begin{NotAbridged}
For instance, when tagging a newly created stack slot, we simulate the memory tag write by writing the used pointer into the stack slot itself.
One special case is the \inlineasm{st2g} instruction that is used to write a memory tag to two adjacent 16-byte granules.
In some cases, the two 4-bit tag writes of \inlineasm{st2g} could happen on a cache-line boundary.
To ensure our upper bound accounts for this, we continue to use a single write, but shift the address by 12-bytes to ensure the emulated \gls{MTE} overhead causes multiple cache-line updates at least as often as \inlineasm{st2g} would have.
\end{NotAbridged}

\textbf{Tag checks} implemented in hardware are likely to have minimal overhead.
We cannot simulate the overhead without substantially inflating the estimate.
However, the tag-check overhead is expected to be negligible and uniformly applied to all memory accesses.
Consequently it is not included in our measurements, and we note that real \gls{MTE}-capable hardware will also induce a small to negligible overhead due to allocation tag lookup and check on memory accesses.

\subsubsection{Results}

Benchmark results and statistics on instrumentation are summarized in \Cref{tbl:spec-overhead}.
The \gls{GM} of the execution-time overhead is \specPerfOverhead{}.
Comparing this to the estimated $7\%$ MTE tag-checking overhead reported by~\cite{lemayCryptographic2021}, this suggests that the \tagProt instrumentation is a significant contributor to the overhead.
The \safeColumn{} column in the table shows the proportion of allocations in bytes that is tagged as safe, but because this can be skewed by large allocations and is based on static information, it does not necessarily correlate with the frequency of \tagProt events at run-time.
We also expect that the \tagProt overhead can be lowered by optimizing it to avoid redundant tag checking, and further by more accurate analysis and by optimizing the use of stack guards.
Nonetheless, even the test that performs worst, 500.perlbench\_r, only reaches an overhead of \specPerfOverheadMax{}, indicating that \IMPLNAME{} performance overhead is efficient (\ref{req:perf}).

\begin{table*}[tp]
	\centering
	\begin{filecontents*}{csv/spec-overhead.csv}
bn,aumean+std,bumean+std,oumean+check,osize,ostack,cbsaf
500.perlbench\_r,201.87 ($\pm$1.29),233.29 ($\pm$1.4),1.166,1.301,1.227,54.985\%
502.gcc\_r,109.7 ($\pm$0.26),130.13 ($\pm$0.65),1.196,1.338,1.309,78.97\%
505.mcf\_r,656.7 ($\pm$0.41),769.61 ($\pm$1.44),1.182,1.143,1.017,4.505\%
519.lbm\_r,453.14 ($\pm$0.39),460.86 ($\pm$2.41),1.027,1.092,1.241,70.079\%
525.x264\_r,140.37 ($\pm$0.29),176.53 ($\pm$0.39),1.268,1.249,1.143,41.544\%
538.imagick\_r,736.29 ($\pm$3.47),763.65 ($\pm$0.36),1.047,1.239,1.129,9.118\%
544.nab\_r,737.19 ($\pm$0.41),765.45 ($\pm$1.15),1.048,1.235,1.145,39.381\%
557.xz\_r,157.22 ($\pm$1.39),183.75 ($\pm$0.23),1.179,1.159,1.366,32.071\%
geo.mean,,,1.136,1.217,1.193,
\end{filecontents*}
\csvreader[%
			before reading=\small,
			tabular=|l|r*{10}{|r}|,
			table head=%
				\hline
				\multirow{2}{*}{Benchmark}
					& \multicolumn{1}{c|}{Baseline}
					& \multicolumn{1}{c|}{Instrumented}
					& \multicolumn{1}{c|}{Safe}
					& \multicolumn{3}{c|}{Overhead}
					\\%
					& time (std.dev)
					& time (std.dev)
					& 
					& time & stack & \texttt{.text}
					\\%
				\hline\hline,
				late after line = \\\hline%
		]{csv/spec-overhead.csv}{}{%
			\csvcoli{}
			& \csvcolii{}
			& \csvcoliii{}
			& \csvcolvii{}{}
			& \csvcoliv{} & \csvcolvi{} & \csvcolv{}
		}%
	\caption{
		Benchmark results using SPEC CPU 2017 benchmarks and the \texttt{ref} workloads.
		\processifversion{NotAbridged}{%
		  The \safeColumn{} column indicates the proportion of the stack (in bytes) that was declared safe by our instrumentation.
		  Overhead figures are normalized to a non-instrumented baseline, and the aggregate overhead (last row) is reported as the \gls{GM}.
		  The stack size estimate is based on summing up the static stack size of all functions in the binary and the binary size increase estimate was calculated by comparing the \texttt{.text} section size of the resulting binaries.
		}
	}%
	\label{tbl:spec-overhead}
\end{table*}

For the binary \texttt{.text} segment size, we see an increase of \specSizeOverhead, \gls{GM}.
The increased size consists of additional code to tag stack slots, and to perform \tagProt.
We observe a stack-size increase of \specMemOverhead (\gls{GM}), excluding the tag memory.
This overhead is due to the need to pad and align tagged allocations to 16-bytes, and because \guarded allocations must be interleaved such that the stack frame is split into disjoint sections with different tags.
The memory use overhead largely depends on how stack memory is used; frequent small allocations quickly accumulate overhead due to the alignment requirements. Indeed, in the benchmarks we notice widely varying memory overheads; from an almost 50\% increase in average stack size for 519.lbm\_r, to a negligible $\sim$1\% increase in 505.mcf\_r. When also accounting for the very low  \safeColumn{} value in the latter test, we can postulate that the 505.mcf\_r code includes large, frequently used but relatively few unsafe stack allocations.

%% file: sections/090.related.tex
\section{Related Work}

Tagged computer architectures emerged in the 1970s to provide data typing and data isolation. 
For instance, the Burroughs B6700 used memory tags to realize typed memory~\cite{lakosImplementing1980}.
In these hardware architectures, type and data were co-located in registers and often also in memory. 
Although this form of fine-grained memory typing disappeared from commercial computing for 50 years, coarse-grained hardware-assisted memory tagging has recently re-emerged in contemporary processor designs such as lowRISC~\cite{bradburyTagged2014}, SPARC processors with the \glsdesc{ADI} feature~\cite{aingaran2015m7}, and AArch64 \gls{MTE} used by this paper~\cite{arm-mte8.5}.
The CHERI capability architecture also uses memory tagging, but only to protect the integrity of the capabilities that then include metadata for checking validity of memory accesses~\cite{watsonCHERI2015}.

Today, Clang \gls{HWASAN}~\cite{serebryanyMemory2018} uses memory tagging to improve performance of \gls{ASAN}~\cite{serebryanyAddressSanitizer2012}.
As discussed in \Cref{sec:background-llvm-memory-tagging}, \gls{HWASAN} does not use \gls{MTE}, but rather enables the use of software-defined tagging with the \gls{TBI} feature.
\Gls{HWASAN} relies on a purely probabilistic tagging scheme for detection of memory errors.
Given the risk of tag forgery, the limited tag size, and the probabilistic approach to tag allocation, \gls{HWASAN} is mainly used as a testing tool, rather than as a run-time protection scheme.

In contrast, LLVM's MemTagSanitizer~\cite{llvm-11-memtagsan} targets post-deployment run-time protection.
Currently, it only supports stack protection, although heap-protection is planned via the Scudo hardened allocator~\cite{llvm-11-scudo}.
While MemTagSanitizer is nominally intended to provide run-time protection, it 
is not designed for the standard adversary model (\Cref{sec:model}) for memory safety.
Specifically, it cannot withstand an adversary that can read tags, and subsequently, inject correctly-tagged pointers.
Moreover, while it prevents most overflows, it relies on random tagging to detect arbitrary writes and use-after-free type errors.
In contrast, we target the standard, more powerful, adversary model.
Nonetheless, MemTagSanitizer could be combined with our work to mitigate errors between unsafe allocations, while using our instrumentation and \tagProt to guarantee the integrity of safe allocations.

The recent HAKC uses both ARM \gls{PA} and \gls{MTE} to realize compartmentalization within the kernel~\cite{mckeePreventing2022}.
\Gls{PA} also covers the \gls{MTE} bits in a pointer, and HAKC relies on it for tag integrity.
HAKC requires the developer to define compartmentalization policies for the kernel and to explicitly define data-ownership transfers between different modules.
Consequently, the overhead also heavily depends on the compartmentalization strategy, ranging from 2\%-\%4 in a single-compartment case, to a linear increase of 14\%-19\% percent per compartment.

We build upon the SafeStackAnalysis introduced by~\cite{kuznetsovCodepointer2014}.
In their work, safe allocations are protected by moving them to a Safe Stack at a randomized location.
It relies on \gls{ASLR} which is not resistant to the standard adversary model that includes full memory disclosure~\cite{serebryanyMemory2018}.
Relying on shadow stacks also changes the system \gls{ABI}, making integration with legacy code difficult.

The more recent Data Guard~\cite{huangTaming2022} introduces an improved static analysis scheme that, similar to our work, can recognize pointers that are safely stored.
While their analysis accuracy is impressive, Data Guard uses Safe Stack for protection, with the shortcomings listed above.

WIT is a \gls{DFI} scheme that uses software-based memory tagging to restrict data-writes to only allowed objects based on a data-flow graph~\cite{akritidisPreventing2008}.
It uses a separate metadata table to store tags and static run-time checks to verify that written memory is tagged with one of the expected values.
Because WIT uses static checks, it cannot, at run-time, distinguish between different valid pointers that can be differentiated by \gls{MTE} address tags.
HDFI~\cite{songHDFI2016} describes RISC-V support for tagging, and similarly to WIT, uses it for \gls{DFI}.
HDFI shows the applicability of tagging to coarse-grained isolation as well as a building block for memory safety solutions such as shadow stacks.
HDFI reports very low overhead ($<2\%$), but uses only a single tag bit, and requires static checks using specific load and store instructions.

%% file: sections/100.discussion.tex
\section{Future work and Conclusion}%
\label{sec:future}

\begin{NotAbridged}
  Languages such as Rust~\cite{matsakisRust2014} can often be combined with unsafe code, weakening the memory-safety guarantees provided by the language.
  Previous work has addressed this with coarse-grained fault-isolation~\cite{lamowskiSandcrust2017}, more fine-grained solutions could offer better performance and flexibility.
  Such language front-ends already use powerful safety analysis for ``safe'' code and can provide this information to the compiler-back end via \gls{IR} safety hint attributes.
  To this end, our implementation of \MTStack adheres to allocation safety hints by ignoring our \gls{IR}-based analysis and treating such allocations as safe, and thus preventing unsafe allocations from corrupting allocations deemed safe by the language front-end.
  In ongoing parallel work, we have prototyped this feature with the LLVM Rustc compiler, and consider this to be a promising direction in general.
\end{NotAbridged}

\begin{NotAbridged}
While our current work is focused on stack-protection, our work is compatible with complementary tagging techniques that mitigate errors on the heap.
We also plan to evaluate our analysis and instrumentation strategy for heap-based allocations and global variables.
\end{NotAbridged}

Due to their scarcity, we do not consider pointers loaded from uninitialized memory, hence, \theAttacker could attempt to exploit a use-before-initialization error to inject unsafe pointer.
Such errors are relatively rare---absent in the top 25 CWEs of 2021~\cite{mitre2021top25}---and can be detected using existing sanitizer such as the LLVM MemorySanitizer~\cite{stepanov2015memorysanitizer}.
As future work, we will nonetheless consider extending the pointer-safety analysis to also verify initialization before use.

Our current work only reasons about pointers in memory to a limited degree, thus limiting the pointers we can assume are safe, and consequently increasing instrumentation overhead.
While more accurate static analysis can improve the situation, not all pointers will be proven safe.
The recently introduced ARM \gls{PA} extension could alleviate this by providing probabilistic integrity guarantees for pointers stored in memory even if they cannot be proven pointer-safe~\cite{qualcommPointer2017,liljestrandPAC2019,liljestrandPACStack2021}.
As \gls{PA} also protects address tags, this could allow the software-based \tagProt to be dropped, consequently offsetting the overhead cause by the  \gls{PA} instrumentation.
Similarly, the memory consumption due to introduced guard pages can be optimized in the compiler by allocation placement so that adjoining allocations in the stack carry different tags---this also enables support for larger guard areas, allowing cases with more linear overflow to be awarded \guarded allocation status.
 
Our work is the first design for deterministic memory safety in a strong adversary model, using an \gls{MTE}-based compartmentalization design, that fully protects stack control-flow data and data that can be proven safe. The first processors with \gls{MTE} were announced very recently~\cite{samsungSamsung2022}. It is reasonable to expect other ARM processor manufacturers to follow suit. Therefore, our work exploring novel ways of using \gls{MTE} is timely.
Our work also provides a clear direction to designers of future memory tagging schemes:  incorporating support for \tagProt (\Cref{sec:design-tag-prot}) into their hardware design can substantially reduce the performance overhead.

%% file: sections/110.acknowledgments.tex
\taggedparagraph{Acknowledgements}

Thomas Nyman and Gilang Hamidy participated in early exploratory work on this topic.
This work is supported in part by Natural Sciences and Engineering Research Council of Canada (RGPIN-2020-04744) and Intel Labs via the Private-AI consortium.